%% file: cluster_computing.tex
\RequirePackage{fix-cm}
\documentclass[twocolumn]{article}          

\usepackage[font=footnotesize]{subfig}
\usepackage{color}
\usepackage{graphicx}
\usepackage{algorithmicx}
\usepackage{algpseudocode}
\usepackage{hyperref} 
\newcommand{\revised}[1]{{\color{black}#1}}

\begin{document}

\title{\revised{Towards Operator-less Data Centers Through Data-Driven, Predictive, Proactive Autonomics}}
 

\author{Alina S\^irbu$^1$        and
        Ozalp Babaoglu $^2$\\
        $^1$Department of Computer Science, University of Pisa, Italy\\
              $^2$Department of Computer Science and Engineering, University of Bologna, Italy
}

\date{}

\maketitle

\begin{abstract}
Continued reliance on human operators for managing data centers is a major impediment for them from ever reaching extreme dimensions.  Large computer systems in general, and data centers in particular, will ultimately be managed using predictive computational and executable models obtained through data-science tools, and at that point, the intervention of humans will be limited to setting high-level goals and policies rather than performing low-level operations. \emph{Data-driven autonomics}, where management and control are based on holistic predictive models that are built and updated using live data, opens one possible path towards limiting the role of operators in data centers.
In this paper, we present a data-science study of a public Google dataset collected in a 12K-node cluster with the goal of building and evaluating predictive models for node failures.   \revised{Our results support the practicality of a data-driven approach by showing the effectiveness of predictive models based on data found in typical data center logs.}  We use BigQuery, the big data SQL platform from the Google Cloud suite, to process massive amounts of data and generate a rich feature set characterizing node state over time. We describe how an ensemble classifier can be built out of many Random Forest classifiers each trained on these features, to predict if nodes will fail in a future 24-hour window. Our evaluation reveals that if we limit false positive rates to 5\%, we can achieve true positive rates between 27\% and 88\%  with precision varying between 50\% and 72\%. \revised{This level of performance allows us to recover large fraction of jobs' executions (by redirecting them to other nodes when a failure of the present node is predicted) that would otherwise have been wasted due to failures.} We discuss the feasibility of including our predictive model as the central component of a data-driven autonomic manager and operating it on-line with live data streams (rather than off-line on data logs). 
All of the scripts used for BigQuery and classification analyses are publicly available on \revised{GitHub}.

\end{abstract}

\section{Introduction}
\label{intro}
Modern data centers are the engines of the Internet that run e-commerce sites, cloud-based services accessed from mobile devices and power the social networks utilized each day by hundreds of millions of users.  Given the pervasiveness of these services in many aspects of our daily lives, continued availability of data centers is critical. And when continued availability is not possible, service degradations and outages need to be foreseen in a timely manner so as to minimize their impact on users. For the most part, current automated data center management tools are limited to low-level infrastructure provisioning, resource allocation, scheduling or monitoring tasks with no predictive capabilities. This leaves the brunt of the problem in detecting and resolving undesired behaviors to armies of operators who continuously monitor streams of data being displayed on monitors.  Even at the highly optimistic rate of 26,000 servers managed per staffer\footnote{Delfina Eberly, Director of Data Center Operations at Facebook, speaking on ``Operations at Scale'' at the \emph{7x24 Exchange 2013 Fall Conference}.}, this situation is not sustainable if data centers are ever to be scaled to extreme dimensions. Applying traditional autonomic computing techniques to large data centers is problematic since their complex system characteristics prohibit building a ``cause-effect'' system model that is essential for closing the control loop.  Furthermore, current autonomic computing technologies are reactive and try to steer the system back to desired states only after undesirable states are actually entered --- they lack predictive capabilities to anticipate undesirable states in advance so that proactive actions can be taken to avoid them in the first place.

If data centers are the engines of the Internet, then data is their fuel and exhaust.  Data centers generate and store vast amounts of data in the form of logs corresponding to various events and errors in the course of their operation.  When these computing infrastructure logs are augmented with numerous other internal and external data channels including power supply, cooling, management actions such as software updates, server additions/removal, configuration parameter changes, network topology modifications, or operator actions to modify electrical wiring or change the physical locations of racks/server/storage devices, data centers become ripe to benefit from  data science.  The grand challenge is to exploit the toolset of modern data science and develop a new generation of autonomics, which we call \revised{\emph{Autonomics 2.0}}, that is \emph{data-driven}, \emph{predictive} and \emph{proactive} based on holistic models that capture a data centre as an \emph{ecosystem} including not only the computer system as such, but also its physical as well as its socio-political environment.

\revised{In this paper we present the results of an initial study of predictive models for node failures in data centers.  Such models will be an integral part of an ``Autonomics 2.0'' architecture that will also include predictive models built from other data sources as well as components to enact proactive control.}  Our study is based on a recent Google dataset containing workload and scheduler events emitted by the Borg cluster management system~\cite{borg} in a cluster of over 12,000 nodes during a one-month period~\cite{googleData,reiss2012b}. 
We employed BigQuery~\cite{bigquery}, a big data tool from the Google Cloud Platform that allows SQL-like queries to be run on massive data, to perform an exploratory feature analysis.  This step generated a large number of features at various levels of aggregation suitable for use in a machine learning classifier. 
The use of BigQuery has allowed us to complete the analysis for large amounts of data
(table sizes up to 12TB containing over 100 billion rows) in reasonable amounts of time.

For the classification study, we employed an ensemble that combines the output of multiple \emph{Random Forests} (RF) classifiers, which themselves are ensembles of Decision Trees.  RF were employed due to their proven suitability in situations were the number of features is large~\cite{rokach2010} and the classes are ``unbalanced''~\cite{khoshgoftaar2007} such that one of the classes consists mainly of ``rare events'' that occur with very low frequency.
Although individual RF were better than other classifiers that were considered in our initial tests, they still exhibited limited performance, which prompted us to pursue an \emph{ensemble approach}. While individual trees in RF are based on subsets of features, we used a combination of \emph{bagging} and data \emph{subsampling} to build the RF ensemble and tailor the methodology to this particular dataset.
Our ensemble classifier was tested on several days from the trace data, resulting in very good performance on some days (up to 88\% true positive rate, TPR, and 5\% false positive rate, FPR), and modest performance on other days (minimum of 27\% TPR at the same 5\% FPR). Precision levels in all cases remained between 50\% and 72\%. We should note that these results are comparable to other failure prediction studies in the field. 

The contributions of our work are severalfold. 
First, we argue that modern data centers can be scaled to extreme dimensions only after eliminating reliance on human operators by adopting a new generation of autonomics that is data-driven and based on holistic predictive models. Towards this goal, we present a failure prediction analysis based on a dataset that has been studied extensively in the literature from other perspectives. 
Next, we propose an ensemble classification methodology tailored to this particular problem where subsampling is combined with
bagging and precision-weighted voting to maximize performance.  Finally, we provide one of the first instances of BigQuery usage in the literature with quantitative evaluation of running times as a function of data size.
Our results show that models with sufficient predictive powers can be built based on data found in typical logs and that they can form the basis of an effective data-driven, predictive and proactive autonomic manager.
\revised{All of the scripts used for BigQuery and classification analysis are publicly available on GitHub~\cite{scripts} under the GNU General Public License.}

The rest of this paper is organized as follows. The next section describes the process of building features from the trace data.  Section~\ref{prediction} describes our classification approach while our prediction results are presented in Section~\ref{results}. Impact on wasted resources is estimated in Section~\ref{impact} while in Section~\ref{discussion} we discuss the issues surrounding the construction of a data-driven autonomic controller based on our predictive model and argue its practicality.  Related work is discussed in Section~\ref{related} while Section~\ref{conclusions} concludes the paper.

\section{Building the feature set with BigQuery}\label{data_analysis}

The workload trace published by Google contains several tables monitoring the status of nodes, jobs and tasks during a period of approximately 29 days for a cluster of 12,453 nodes. This includes task events (over 100 million records, 17GB uncompressed), which follow the state evolution for each task, and task usage logs (over 1 billion records, 178GB uncompressed), which report the amount of resources per task at approximately 5 minute intervals. We have used the data to compute the overall load and status of different cluster nodes at 5 minute intervals. 
This resulted in a time series for each node and feature that spans the entire trace (periods when the node was ``up''). 
We then proceeded to obtain several features by aggregating measures in the original data. Due to the size of the dataset, this aggregation analysis was performed using BigQuery on the trace data directly from Google Cloud Storage. We used the \emph{bq} command line tool for the entire analysis, and our scripts are available online through our Web site~\cite{scripts}.

From task events, we obtained several time series for each node with a time resolution of 5 minutes. A total of 7 features were extracted, which count the number of tasks currently \emph{running}, the number of tasks that have \emph{started} in the last 5 minutes and those that have \emph{finished} with different exit statuses --- \emph{evicted}, \emph{failed}, \emph{finished normally}, \emph{killed} or \emph{lost}.  From task usage data, we obtained 5 additional features (again at 5-minute intervals) measuring the load at node level in terms of: \emph{CPU}, \emph{memory}, \emph{disk time}, \emph{cycles per instruction} (CPI) and \emph{memory accesses per instruction} (MAI). 
This resulted in a total of 12 \emph{basic features} that were extracted.
For each feature, at each time step we consider the previous 6 time windows (corresponding to the 
node status during the last 30 minutes) obtaining 72 features in total (12 basic features $\times$ 6 time windows). 

The procedure for obtaining the basic features was extremely fast on the BigQuery platform.  For task counts, we started with constructing a table of running tasks, where each row corresponds to one task and includes its start time, end time, end status and the bide it was running on. Starting from this table, we could obtain the time series for each feature for each node, requiring
between 139 and 939 seconds on BigQuery per feature (one separate table per feature was obtained). The features related to node load were computed by summing over all tasks running on a node in each time window, requiring
between 3585 and 9096 seconds on BigQuery per feature. The increased execution time is due to the increased table sizes (over 1 billion rows).
We then performed a JOIN of all above tables to combine the basic features into a single table with 104,197,215 rows (occupying 7GB).
For this analysis, our experience allows us to judge BigQuery as being extremely fast; an equivalent computation would have taken months to perform on a regular PC.

\begin{table}[t]
\centering
\begin{tabular}{|c||c|c|}
\hline
Aggregation &Average, SD, CV &Correlation \\
\hline
1h&\emph{166 (all features)}&45(6.5)\\
12h&\emph{864 (all features)}&258.8(89.1)\\
24h&284.6(86.6)&395.6(78.9)\\
48h&593.6(399.2)&987.2(590)\\
72h&726.6(411.5)&1055.47(265.23)\\
96h&739.4(319.4)&1489.2(805.9)\\
\hline
\end{tabular}
\caption{Running times required by BigQuery for obtaining features aggregated over different time windows, for two aggregation types: computing \emph{averages}, \emph{standard deviation} (SD) and \emph{coefficient of variation} (CV) versus computing \emph{correlations}. For 1h and 12h windows, average, SD and CV were computed for all features in a single query. For all other cases, the mean (and standard deviation) of the required times per feature are shown.  }
\label{table_times}
\end{table}

A second level of aggregation meant looking at features over longer time windows rather than just the last 5 minutes. At each time step, 3 different statistics --- averages, standard deviations and coefficients of variation --- were computed for each basic feature obtained at the previous step. This was motivated by the suspicion that not only feature values but also their deviation from the mean could be important in understanding system behavior. Six different running windows of sizes 1, 12, 24, 48, 72 and 96 hours were used to capture behavior at various time resolutions. This resulted in 216 additional features (3 statistics $\times$ 12 features $\times$ 6 window sizes).

In order to generate these aggregated features, a set of intermediate tables were used. For each time point, these tables consisted of the entire set of data points to be averaged. For instance, for 1-hour averages, the table would contain a set of 6 values for each feature and for each time point, showing the evolution of the system over the past hour. While generating these tables was not time consuming (requiring between 197 and 960 seconds), their sizes were quite impressive: ranging from 143 GB (over 1 billion rows) for 1 hour up to 12.5 TB (over 100 billion rows) in the case of 96-hour window. Processing these tables to obtain the aggregated features of interest required significant resources and would not have been possible without BigQuery. Even then, direct queries using a single \textsc{group by} operation to obtain all 216 features was not possible, requiring only one basic feature to be handled at a time and combining the results into a single table at the end. Table~\ref{table_times} shows statistics over the time required to obtain one feature for the different window sizes.

Although independent feature values are important, another criterion that could be important for prediction is the relations that exist between different measures. Correlation between features is one such measure, with different correlation values indicating changes in system behavior. Hence we introduced a third level of aggregation of the data by computing correlations between a chosen set of feature pairs, again over various window sizes (1 to 96 hours as before). We chose 7 features to analyze: number of running, started and failed jobs together with CPU, memory, disk time and CPI. By computing correlations between all possible pairings of the 7 features, we obtained a total of 21
correlation values for each window size. This introduces 126 additional features to our dataset. The BigQuery analysis started from the same intermediate tables as before and computed correlations for one feature pair at a time. As can be seen in Table~\ref{table_times}, this step was more time consuming, requiring greater time than the previous aggregation step, yet still remains manageable considering the size of the data. The amount of data processed for these queries ranged from 49.6GB (per feature pair for 1-hour windows) to 4.33TB (per feature pair for 96-hour windows), resulting in a higher processing cost (5 USD per TB processed). Yet again, a similar analysis would not have been possible without the BigQuery platform.

The Google trace also reports node events. These are scheduler events corresponding to nodes being added or removed from the pool of resources. Of particular interest are \textsc{remove} events, which can be due to two causes: node failures or software updates. The goal of this work is to predict \textsc{remove} events due to \emph{node failures}, so the two causes have to be distinguished. Prompted by our discussions, publishers of the Google trace investigated the best way to perform this distinction and suggested to look at the length of time that nodes remain down --- the time from the \textsc{remove} event of interest to the next \textsc{add} event for the same node. If this ``down time'' is large, then we can assume that the \textsc{remove} event was due to a node failure, while if it is small, the node was most likely removed for a software update. To ensure that an event considered to be a failure is indeed a real failure, we used a relatively-long ``down time'' threshold of 2 hours, which is greater than the time required for a typical software update.
Based on this threshold, out of a total of 8,957 \textsc{remove} events, 2,298 were considered failures, and were the target of our predictive study. For the rest of the events, for which we cannot be sure of the cause, the data points in the preceding 24-hour window were removed completely from the dataset. An alternative would have been considering them part of the \textsc{safe} class, however this might not be true for some of the points. Thus, removing them completely ensures that all data labeled as \textsc{safe} are in fact \textsc{safe}.

To the above features based mostly on load measures, we added two new features: the \emph{up time} for each node (time since the last corresponding \textsc{add} event) and
number of \textsc{remove} events for the entire cluster within the last hour.
This resulted in a total of 416 features for 104,197,215 data points (almost 300GB of processed data).


\section{Classification approach}\label{prediction}

The features obtained in the previous section were used for classification with the \emph{Random Forest} (RF) classifier. The data points were separated into two classes: \textsc{safe} (negatives) and \textsc{fail} (positives). To do this, for each data point (corresponding to one node at a given time) we computed \emph{time\_to\_remove} as the time to the next \textsc{remove} event. Then, all points with \emph{time\_to\_remove} less than 24 hours were assigned to the class \textsc{fail} while all others were assigned to the class \textsc{safe}.
We extracted all the \textsc{fail} data points corresponding to real failures (108,365 data points) together with a 
subset of the \textsc{safe} class, corresponding to 0.5\% of the total by random subsampling (544,985 points after subsampling). 
We used this procedure to deal with the fact that the \textsc{safe} class is much larger than the \textsc{fail} class and classifiers have difficulty learning patterns from very imbalanced datasets. Subsampling is one way of reducing the extent of this imbalance~\cite{galar2012}. Even after this subsampling procedure, negatives are about five times the number of positives.
These 653,350 data points (\textsc{safe} plus \textsc{fail}) formed the basis of our predictive study.

Given the large number of features, some might be more useful than others, hence we explored two types of feature selection mechanisms. One was \emph{principal component analysis} (PCA), using the original features to build a set of principal components --- additional features that account for most of the variability in the data. Then one can use only the top principal components for classification, since those should contain the most important information. We trained classifiers with an increasing number of principal components, however the performance obtained was not better than using the original features. A second mechanism was to filter the original features based on their correlation to the time to the next failure event (\emph{time\_to\_remove} above). Correlations were in the interval $[-0.3,0.45]$, and we used only those features with absolute correlation larger than a threshold. We found that the best performance was obtained with a null threshold, which means again using all features. Hence, our attempts to reduce the feature set did not produce better results that the RF trained directly on the original features. One reason for this may be the fact that the RF itself performs feature selection when training the Decision Trees. It appears that the RF mechanism performs better in this case than correlation-based filtering or PCA. 

\begin{figure}[!t]
\centering
\includegraphics[width=3in]{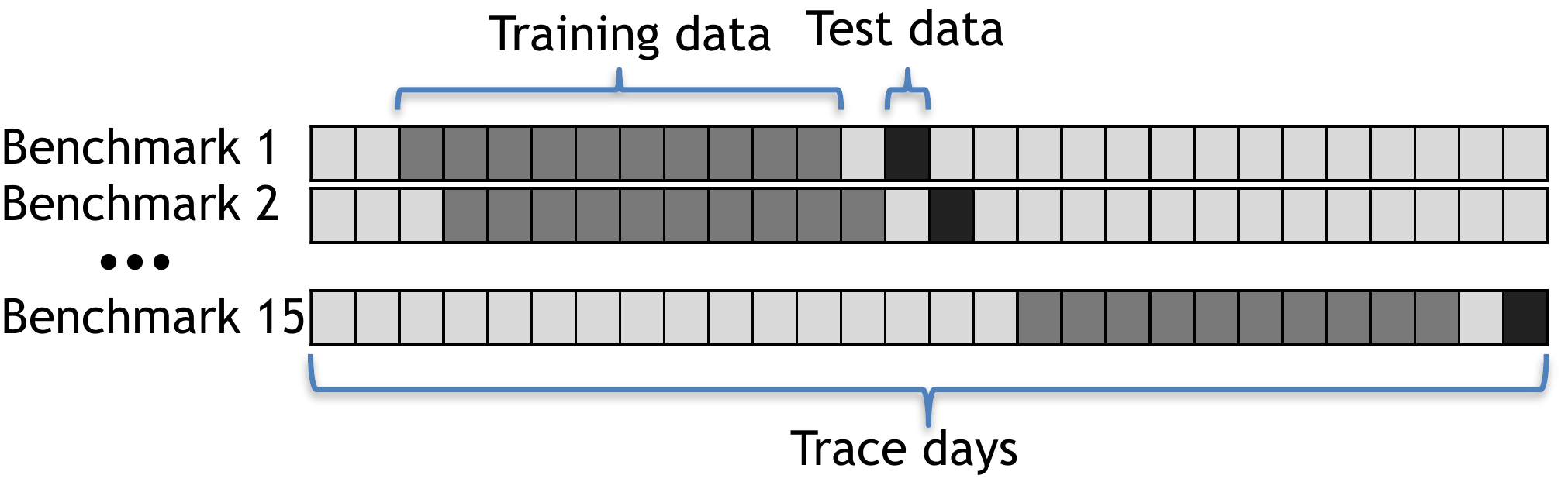}
\caption{Cross validation approach: forward-in-time testing. Ten days were used for training and one day for testing. A set of 15 benchmarks (train/test pairs) were obtained by sliding the train/test window over the 29-day trace.}
\label{fig_xval}
\end{figure}

To evaluate the performance of our approach, we employed cross validation. Given the procedure we used to define the two classes, there are multiple data points corresponding to the same failure (data over 24 hours with 5 minutes resolution). Since some of these data points are very similar, choosing the \emph{train} and \emph{test} data cannot be done by selecting random subsets. While random selection may give extremely good prediction results, it is not realistic since we would be using test data which is too similar to the training data. This is why we opted for a time-based separation of train and test data. We considered basing the training on data over a 10-day window, followed by testing based on data over the next day with no overlap with the training data. Hence, the test day started 24 hours after the last training data point. The first two days were omitted in order to decrease the effect on aggregated features. In this manner, fifteen train/test pairs were obtained and used as benchmarks to evaluate our analysis (see Fig.~\ref{fig_xval}). This forward-in-time cross validation procedure ensures that classification performance is realistic and not an artifact of the structure of the data. Also, it mimics the way failure prediction would be applied in a live data center, where every day a model could be trained on past data to predict future failures.

Given that many points from the \textsc{fail} class are very similar, which is not the case for the \textsc{safe} class due to initial subsampling, the information in the \textsc{safe} class is still overwhelmingly large. This prompted us to further subsample the negative class in order to obtain the training data. This was performed in such a way that the ratio between \textsc{safe} and \textsc{fail} data points is equal to a parameter \emph{fsafe}. 
We varied this parameter with the values  $\{0.25,0.5,1,2,3,4\}$ while using all of the data points from the positive class so as not to miss any useful information. 
This applied only for training data: for testing we always used all data from both the negative and positive classes (out of the base dataset of 653,350 points).
We also used RF of different sizes, with the number of Decision Trees varying from 2 to 15 
with a step of 1 (resulting in 14 different values).

As we will discuss in the next section, the performance of the individual classifiers, while better than random, was judged to be not satisfactory. Hence we opted for an ensemble method, which builds a series of classifiers and then selects and combines them to provide the final classification. Ensembles can enhance the power of low performing individual classifiers \cite{rokach2010}, especially if these are diverse~\cite{kuncheva2000,shipp2002}: if they give false answers on different data points (independent errors), then combining their knowledge can improve accuracy. To create diverse classifiers, one can vary model parameters but also train them with different data (known as the \emph{bagging} method \cite{rokach2010}). Bagging matches very well with subsampling to overcome the rare events issue, and it has been shown to be effective for the class-imbalance problem~\cite{galar2012}. Hence, we adopt a similar approach to build our individual classifiers. Every time a new classifier is trained, a new training dataset is built by considering all the data points in the positive class and a random subset of the negative class. As described earlier, the size of this subset is defined by the \emph{fsafe} parameter. By varying the value of \emph{fsafe} and number of trees in the RF algorithm, we created diverse classifiers. The following algorithm details the procedure of building the individual classifiers in the ensemble.

\vspace{1em}
{\small
\begin{algorithmic}
\Require \emph{train\_pos}, \emph{train\_neg}, \emph{Shuffle}(), \emph{Train}()
\State \emph{fsafe} $\gets \{0.25,0.5,1,2,3,4\}$
\State \emph{tree\_count} $\gets \{2..15\}$
\State \emph{classifiers} $\gets \{\}$
\State \emph{start} $\gets 0$
\ForAll {\emph{fs} $\in$ \emph{fsafe}}
\ForAll {\emph{tc} $\in$ \emph{tree\_count}}
\State \emph{end} $\gets$ \emph{start} $+$ \emph{fs} $*$ $|$\emph{train\_pos}$|$
\If {\emph{end} $\geq$ $|$\emph{train\_neg}$|$}
    \State \emph{start} $\gets 0$
    \State \emph{end} $\gets$ \emph{start} $+$ \emph{fs} $*$ $|$\emph{train\_pos}$|$
    \State \emph{Shuffle}(\emph{train\_neg})
\EndIf
\State \emph{train\_data} $\gets$ \emph{train\_pos} $+$ \emph{train\_neg}$[$\emph{start}: \emph{end}$]$
\State \emph{classifier} $\gets$ \emph{Train}(\emph{train\_data}, \emph{tc})
\State \emph{append}(\emph{classifiers}, \emph{classifier})
\State \emph{start} $\gets$ \emph{end}
\EndFor
\EndFor
\end{algorithmic}
}
\vspace{1em}

We repeated this procedure 5 times, resulting in 5 classifiers for each combination of the parameters \emph{fsafe} and RF size. This resulted in a total of 420 RF in the ensemble (5 repetitions $\times$ 6 \emph{fsafe} values $\times$ 14 RF sizes).

Once the pool of classifiers is obtained, a combining strategy has to be used. Most existing approaches use the majority vote rule --- each classifier votes on the class and the majority class becomes the final decision~\cite{rokach2010}. Alternatively, a weighted vote can be used, and we opted for \emph{precision-weighted voting}. For most existing methods, weights correspond to the accuracy of each classifier on training data~\cite{opitz1996}. In our case, performance on training data is close to perfect and accuracy is generally high, which is why we use precision on a subset of the test data. Specifically, we divide the test data into two halves: an \emph{individual test} data set and an \emph{ensemble test} data set. The former is used to evaluate the precision of individual classifiers and obtain a weight for their vote. The latter provides the final evaluation of the ensemble. All data corresponding to the test day was used, with no subsampling. Table \ref{table_datapoints} shows the number of data points used for each benchmark for training and testing. While the parameter \emph{fsafe} controlled the ratio \textsc{safe}/\textsc{fail} during training, \textsc{fail} instances were much less frequent during testing, varying between 13\% and 36\% of the number of \textsc{safe} instances.

\begin{table}[h]
\centering
\begin{tabular}{|c||c||c|c||c|c|}
\hline
&Train&\multicolumn{2}{|c||}{Individual Test}&\multicolumn{2}{|c|}{Ensemble Test}\\
\hline
Benchmark&\textsc{fail}&\textsc{fail}&\textsc{safe}&\textsc{fail}&\textsc{safe}\\
\hline \hline
\input{datapoints.txt}

\hline
\end{tabular}
\caption{Size of training and testing datasets. For training data, the number of \textsc{safe} data points is the number of \textsc{fail} multiplied by the \emph{fsafe} parameter at each run. }
\label{table_datapoints}
\end{table}

To perform precision-weighted voting, we first applied each RF $i$ obtained above to the \emph{individual test data} and computed their precision $p_i$ as the fraction of points labeled \textsc{fail} that were actually failures.  In other words, \emph{precision} is the probability that an instance labeled as a failure is actually a real failure, which is why we decided to use this as a weight. Then we applied each RF to the \emph{ensemble test data}. For each data point $j$ in this set, each RF provided a classification $o_i^j$ (either 0 or 1 corresponding to \textsc{safe} or \textsc{fail}, respectively).  The classification of the ensemble (the whole set of RF) was then computed as a \emph{continuous score} 
\begin{equation}
s_j=\sum_i{o_i^j p_i}
\end{equation}
by summing individual answers weighted by their precision. Finally, these were normalized by the highest score in the data
\begin{equation}\label{eq_score}
s_j'=\frac{s_j}{max_j(s_j)}
\end{equation}
The resulting score $s_j'$ is proportional to the likelihood that a data point is in the \textsc{fail} class --- the higher the score, the more certain we are that we have an actual failure. The following algorithm outlines the procedure of obtaining the final ensemble classification scores.

\vspace{1em}
{\small
\begin{algorithmic}
\Require \emph{classifiers}, \emph{individual\_test}, \emph{ensemble\_test}
\Require \emph{Precision}(), \emph{Classify}()
\State \emph{classification\_scores} $\gets \{\}$
\State \emph{weights} $\gets \{\}$
\ForAll {\emph{c} $\in$ \emph{classifiers}}
\State \emph{w} $\gets$ \emph{Precision}(\emph{c}, \emph{individual\_test})
\State \emph{weights}[\emph{c}] $\gets$ \emph{w}
\EndFor
\ForAll {\emph{d} $\in$ \emph{ensemble\_test}}
\State  \emph{score} $\gets 0$
\ForAll {\emph{c} $\in$ \emph{classifiers}}
\State \emph{score} $\gets$ \emph{score}$+$\emph{weights}[\emph{c}]$*$\emph{Classify}(\emph{c}, \emph{d})
\EndFor
\State \emph{append}(\emph{classification\_scores}, \emph{score})
\EndFor
\State \emph{max} $\gets$ \emph{Max}(\emph{classification\_scores})
\ForAll {\emph{s} $\in$ \emph{classification\_scores}}
\State \emph{s} $\gets$ \emph{s}/\emph{max}
\EndFor
\end{algorithmic}
}
\vspace{1em}

 It assumes that the set of classifiers is available (\emph{classifiers}), together with the two test data sets (\emph{individual\_test} and \emph{ensemble\_test}) and  procedures to compute precision of a classifier on a dataset (\emph{Precision}()) and to apply a classifier to a data point (\emph{Classify}() which returns 0 for \textsc{safe} and 1 for \textsc{fail}).

\begin{figure}[!b]
\centering
\includegraphics[width=3in]{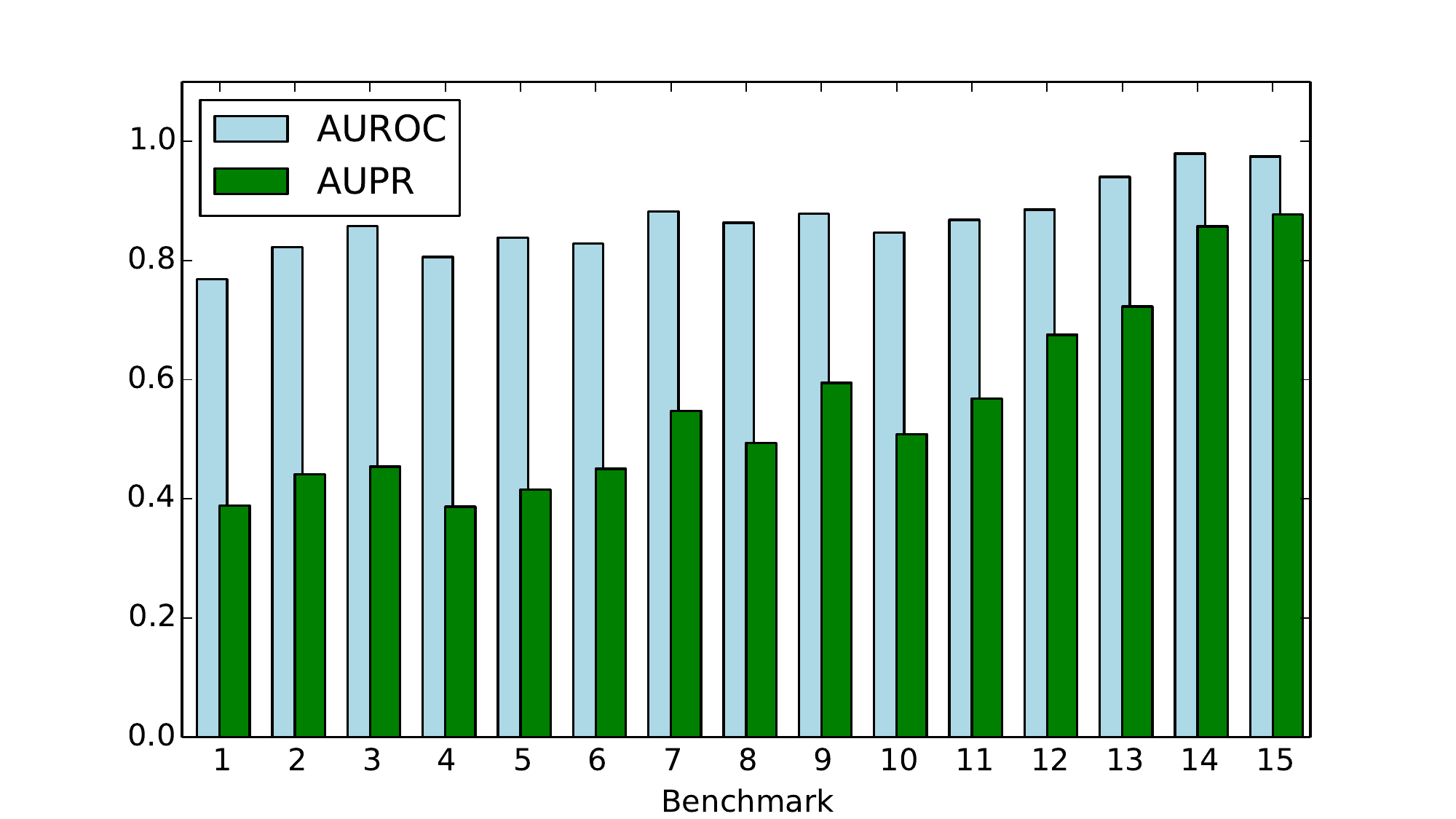}
\caption{AUROC and AUPR on \emph{ensemble test data}.}
\label{fig_au_all}
\end{figure}

\section{Classification results}\label{results}

The ensemble classifier was applied to all 15 benchmark datasets. 
Training was done on an iMac with 3.06GHz Intel Core 2 Duo processor and 8GB of 1067MHz DDR3 memory running OSX 10.9.3. Training of the entire ensemble took between 7 and 9 hours for each benchmark.

Given that the result of the classification is a continuous score (Equation \ref{eq_score}), and not a discrete label, evaluation was based on the \emph{Receiver Operating Characteristic} (ROC) and \emph{Precision-Recall} (PR) curves. A class can be obtained for a data point $j$ from the score $s_j'$ by using a threshold $s^*$.
A data point is considered to be in the \textsc{fail} class if $s_j'\geq s^*$.
The smaller $s^*$, the more instances are classified as failures. Thus, by decreasing $s^*$ the number of true positives grows but so do the false positives. Similarly, at different threshold values, a certain precision is obtained.
The ROC curve plots the True Positive Rate (TPR) versus the False Positive Rate (FPR) of the classifier as the threshold is varied.
Similarly, The PR curve displays the \emph{precision} versus \emph{recall} (equal to TPR or Sensitivity). It is common to evaluate a classifier by computing the \emph{area under ROC} (AUROC) and \emph{area under PR} (AUPR) curves, which can range from 0 to 1. AUROC values greater than 0.5 correspond to classifiers that perform better than random guesses, while AUPR represents an average classification precision, so, again, the higher the better. AUROC and AUPR do not depend on the relative distribution of the two classes, so they are particularly suitable for class-imbalance problems such as the one at hand.

Fig. \ref{fig_au_all} shows AUROC and AUPR values obtained for all datasets, evaluated on the \emph{ensemble test} data.  For all benchmarks, AUROC values are very good, over 0.75 and up to 0.97. AUPR ranges between 0.38 and 0.87. Performance appears to increase, especially in terms of precision, towards the end of the trace. Lower performance that is observed for the first two benchmarks could be due to the fact that some of the aggregated features (those over 3 or 4 days) are computed with incomplete data at the beginning.

\begin{figure*}[!t]
\centering
\subfloat[Worst case (Benchmark 4)]{\includegraphics[width=5in]{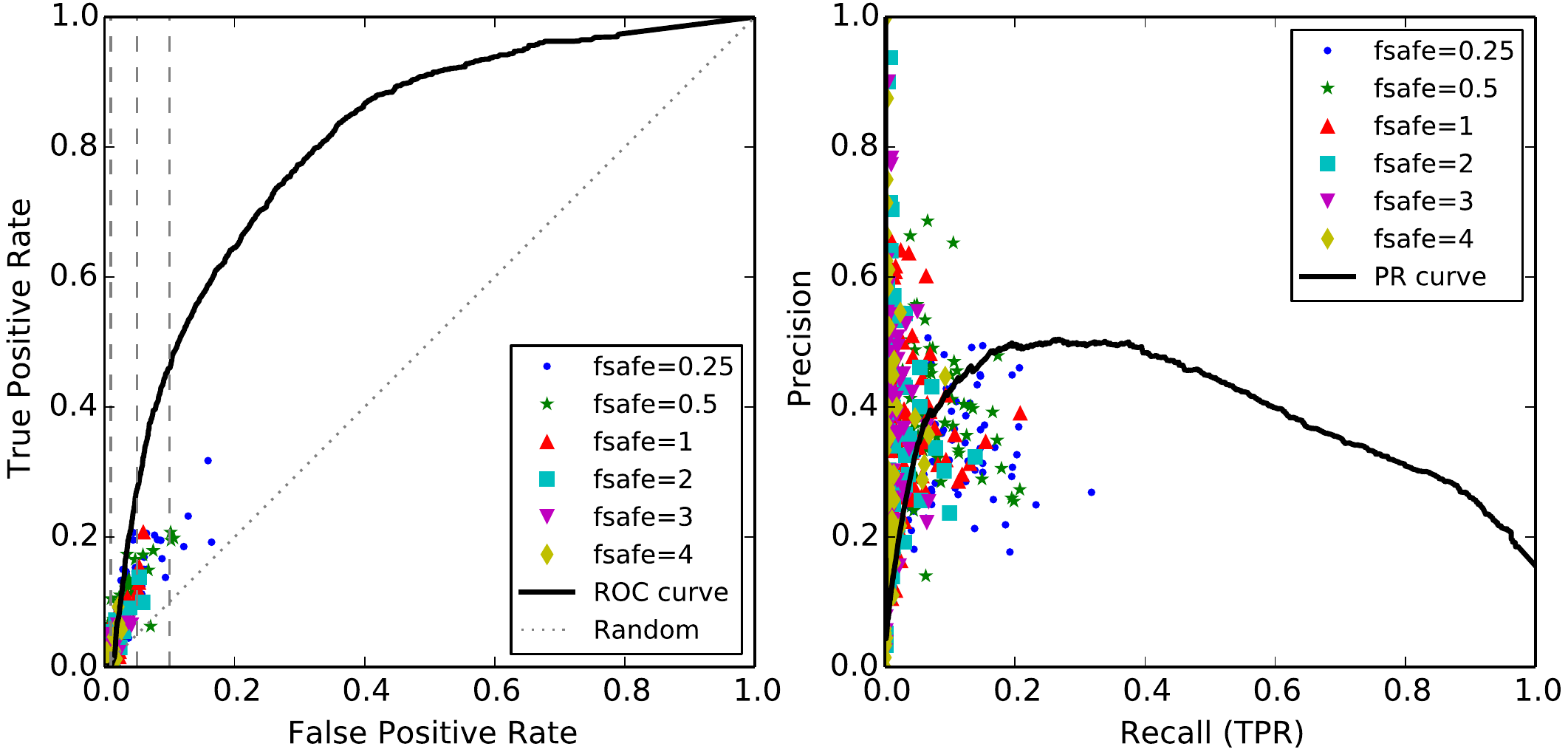}%
\label{fig_curves5}}
\\
\subfloat[Best case (Benchmark 14)]{\includegraphics[width=5in]{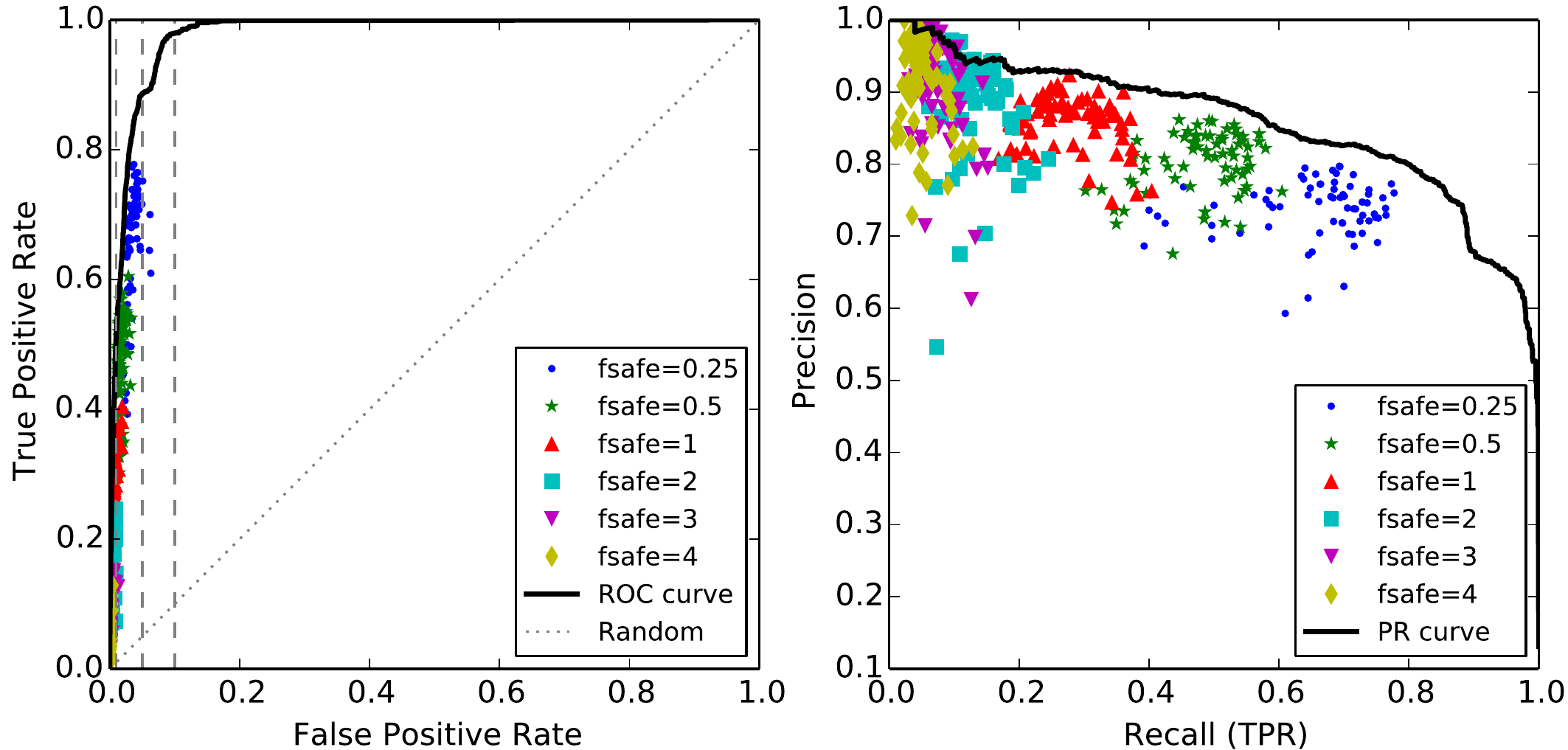}%
\label{fig_curves15}}
\caption{ROC and PR curves for worst and best performance across the 15 benchmarks (4 and 14, respectively). The vertical lines correspond to FPR of 1\%, 5\% and 10\%.
Note that parameter \emph{fsafe} controls the ratio of \textsc{safe} to \textsc{fail} data in the training datasets.}
\label{fig_curves}
\end{figure*}

To evaluate the effect of the different parameters and the ensemble approach, Fig.~\ref{fig_curves} displays the ROC and PR curves for the benchmarks that result in the worst and best results (4 and 14, respectively). Performance of the individual classifiers in the ensemble are also displayed (as points in the ROC/PR space since their answer is categorical). We can see that individual classifiers result in very low FPR which is very important in predicting failures.
Yet, in many cases, the TPR values are also very low. 
This means that most test data is classified as \textsc{safe} and very few failures are actually identified. 
TPR appears to increase when the \emph{fsafe} parameter decreases, but at the expense of the FPR and Precision. The plots show quantitatively the clear dependence between the three plotted measures and \emph{fsafe} values. As the amount of \textsc{safe} training data decreases, the classifiers become less stringent and can identify more failures, which is an important result for this class-imbalance problem. Also, the plot shows clearly that individual classifiers obtained with different values for \emph{fsafe} are diverse, which is critical for obtaining good ensemble performance. 

In general, the points corresponding to the \emph{individual classifiers} are below the ROC and PR curves describing the performance of the \emph{ensemble}. This proves that the ensemble method is better than the individual classifiers for this problem, which can be also due to their diversity. Some exceptions do appear (points above the solid lines), however for very low TPR (under 0.2) so in an area of the ROC/PR space that is not interesting from our point of view. We are interested in maximizing the TPR while keeping the FPR at bay.  At a FPR of 5\%, which means few false alarms, the two examples from Fig.~\ref{fig_curves} display TPR values of 0.272 (worst case) and 0.886 (best case), corresponding to precision values of 0.502 and 0.728 respectively. This is much better than individual classifiers at this level, both in terms of precision and TPR. For failure prediction, this means that between 27.2\% and 88.6\% of failures are identified as such, while from all instances labeled as failures, between 50.2\% and 72.8\% are actual failures.

\revised{According to our classification strategy, a node would be considered to be in \textsc{safe} state whether it fails in 2 days or in 2 weeks. Similarly, it is considered to be in  \textsc{fail} state whether it fails in 10 minutes or within 23 hours. Obviously the two situations are very different and the impact of misclassification varies depending on the time to the next failure. In earlier work \cite{Sirbu2015}, we have shown that misclassifications of time points in the \textsc{fail} class (false negatives) are in general far from the time of failure, while  misclassifications of the \textsc{safe} class (false positives or false alarms) are, on average, closer to the failure instant than are true negatives (correct classification of \textsc{safe} time points). This implies that the impact of misclassification is reduced. }


\section{Impact on running tasks}\label{impact}

\revised{
When a node fails, all tasks running on that node are \emph{interrupted}. Thus, resources (e.g., CPU time) that had been consumed by interrupted tasks are wasted.  To study the extent of this wastage, we estimated the number of interrupted tasks and the corresponding CPU-hours wasted due to failures in our data.  The study was limited to the period of the trace corresponding to \emph{ensemble test data} in all 15 benchmarks representing 180 hours of data and containing 668 node failures. For this period, a simple count of tasks evicted or killed in close vicinity of the failure results in a total of 5,488 interrupted tasks, corresponding to over 31,393 CPU-hours being wasted. 

The Google dataset contains data from different tasks that use resources very differently. Some tasks correspond to latency-sensitive, long-running, revenue-generating production services that respond to user queries. If these tasks are interrupted, only the latest queries will be affected, so most of the CPU time they used is not actually wasted. Other tasks are from non-production batch jobs, and typically return a value at the very end, so their interruption results in all of the CPU time they used being wasted. The Google dataset includes the \emph{scheduling class} of a task to distinguish between production services (higher scheduling classes) and non-production batch jobs (lower scheduling classes). Table~\ref{table_baseline} indicates the distribution of the \emph{interrupted} tasks into the different scheduling classes. We can see that most interrupted tasks are from scheduling class 0 (non-production).
However, most of the wasted CPU time corresponds to higher scheduling classes since these jobs are long running.

\begin{table}[h]
\centering
\begin{tabular}{|c||c|c|c|}
\hline
Class&Interrupted&Recovered&Redirected\\
\hline
     0 &2,260(2,675)&1,863(1,538)&27,908(6,550)\\
     1  &851(1,748) & 387(351) & 4,796(2,294)\\
     2    &   1,632(16,663) &     667(2,035) & 7,586(5,615)\\
     3      &   745(10,305) &     185(1,269)&    791(3584)\\
     \hline\hline
  Total  &    5,488(31,393) &   3,102(5,194) &41,081(18,044)\\
\hline
\end{tabular}
\caption{\revised{Number of tasks in each scheduling class that are \emph{interrupted} by node failures, together with those \emph{recovered} and \emph{redirected} with perfect prediction.
CPU-hours used by the tasks in each category are shown in parentheses (wasted, recovered and redirected CPU-hours).} }
\label{table_baseline}
\end{table}

Predicting failures in advance can help reduce resource wastage by modifying the resource allocation decisions dynamically. This is in itself an extensive research area, but a simple procedure could be to \emph{quarantine} nodes that are predicted to fail by not submitting any new jobs to them for some period of time. If consecutive failure alarms continue to appear, the quarantine period is extended until either the alarms stop or the node fails. In the following, we simulate such an approach. For better precision, we quarantine a node only if two consecutive time points are classified as \textsc{FAIL}.

\begin{figure*}
\centering
\subfloat[Tasks recovered]{\includegraphics[width=2.7in]{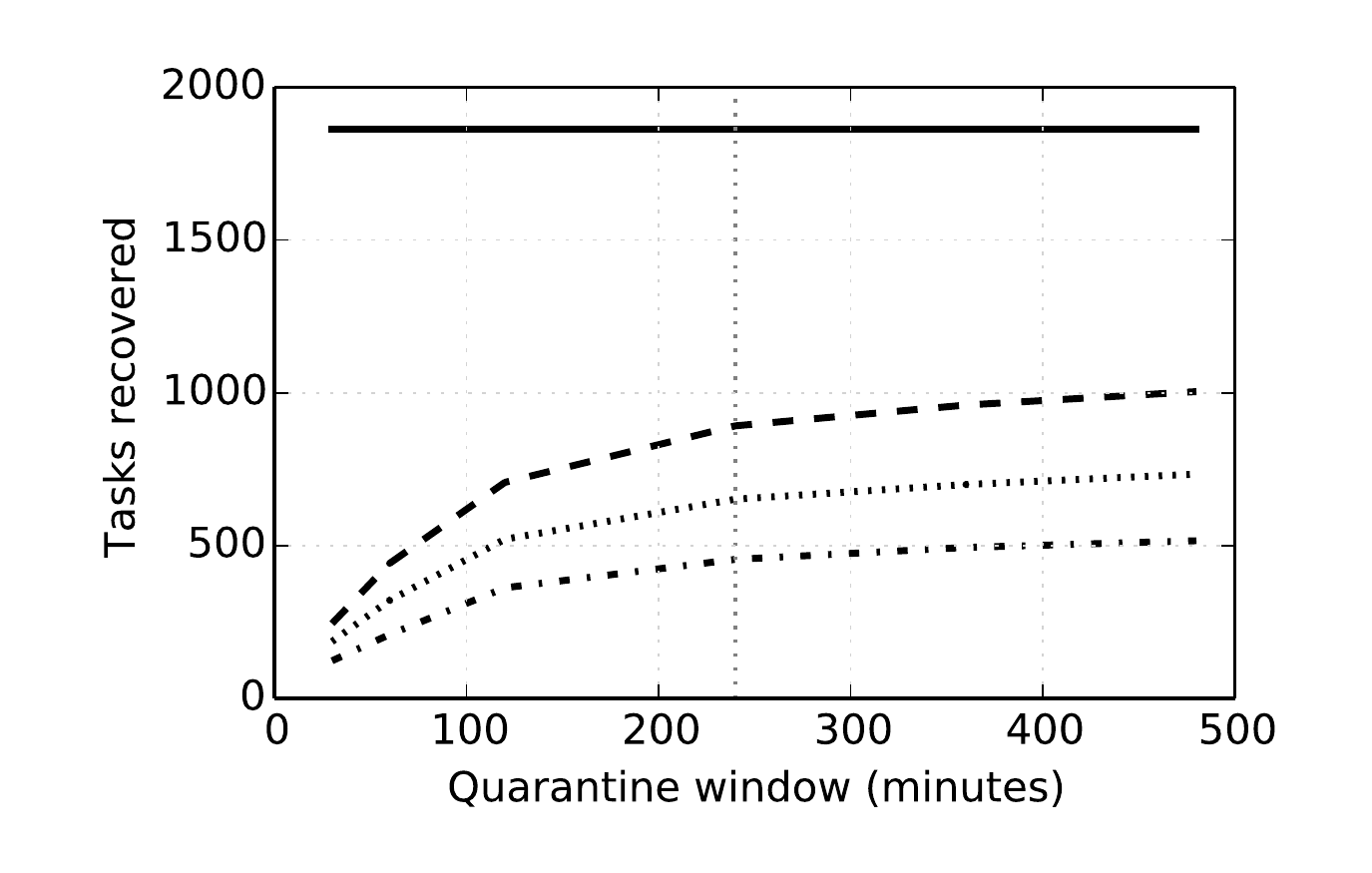}%
\label{fig_ttr5_a}}
\subfloat[CPU time recovered]{\includegraphics[width=2.7in]{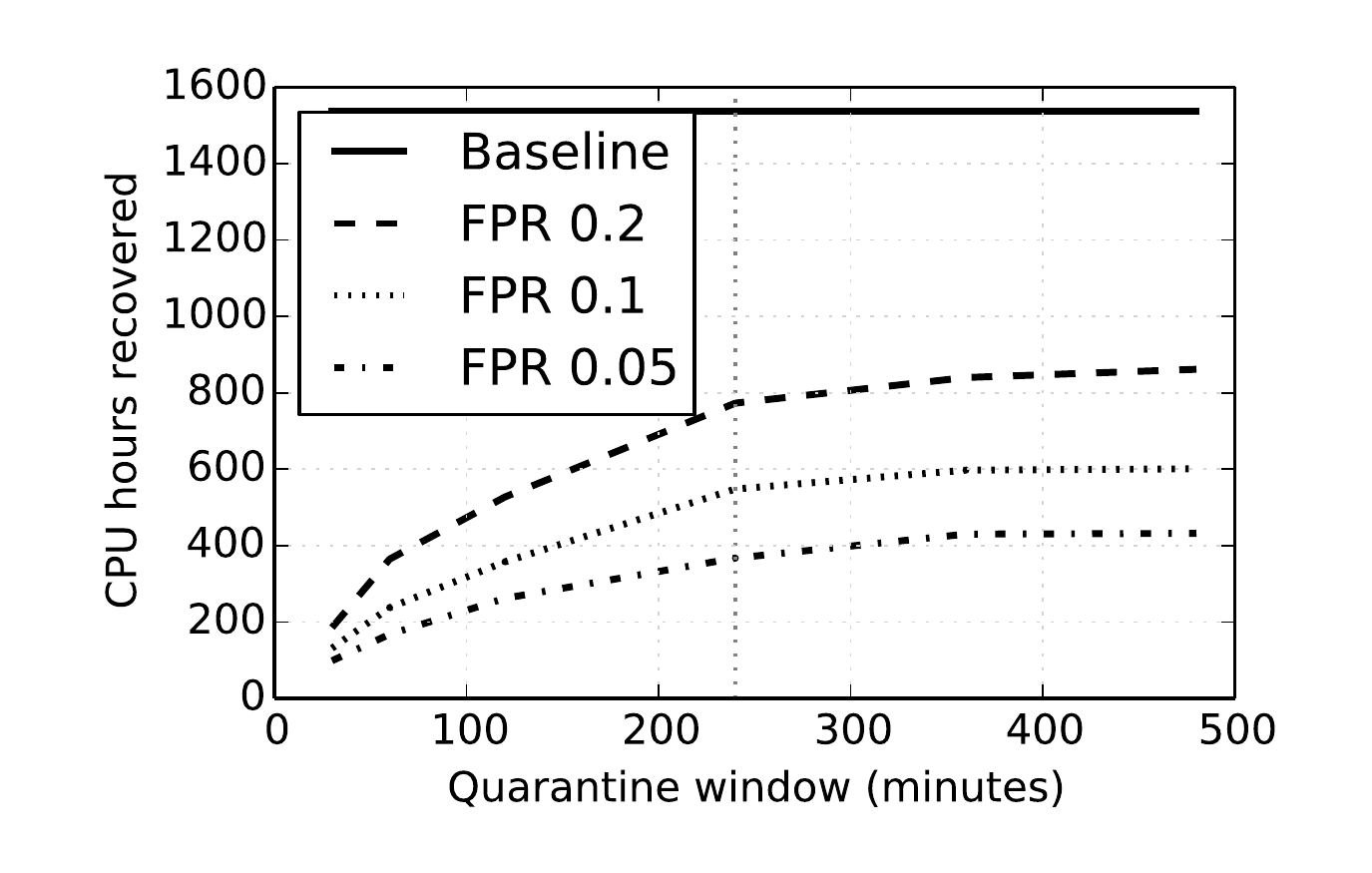}%
\label{fig_ttr5_b}}
\\
\subfloat[Tasks redirected]{\includegraphics[width=2.7in]{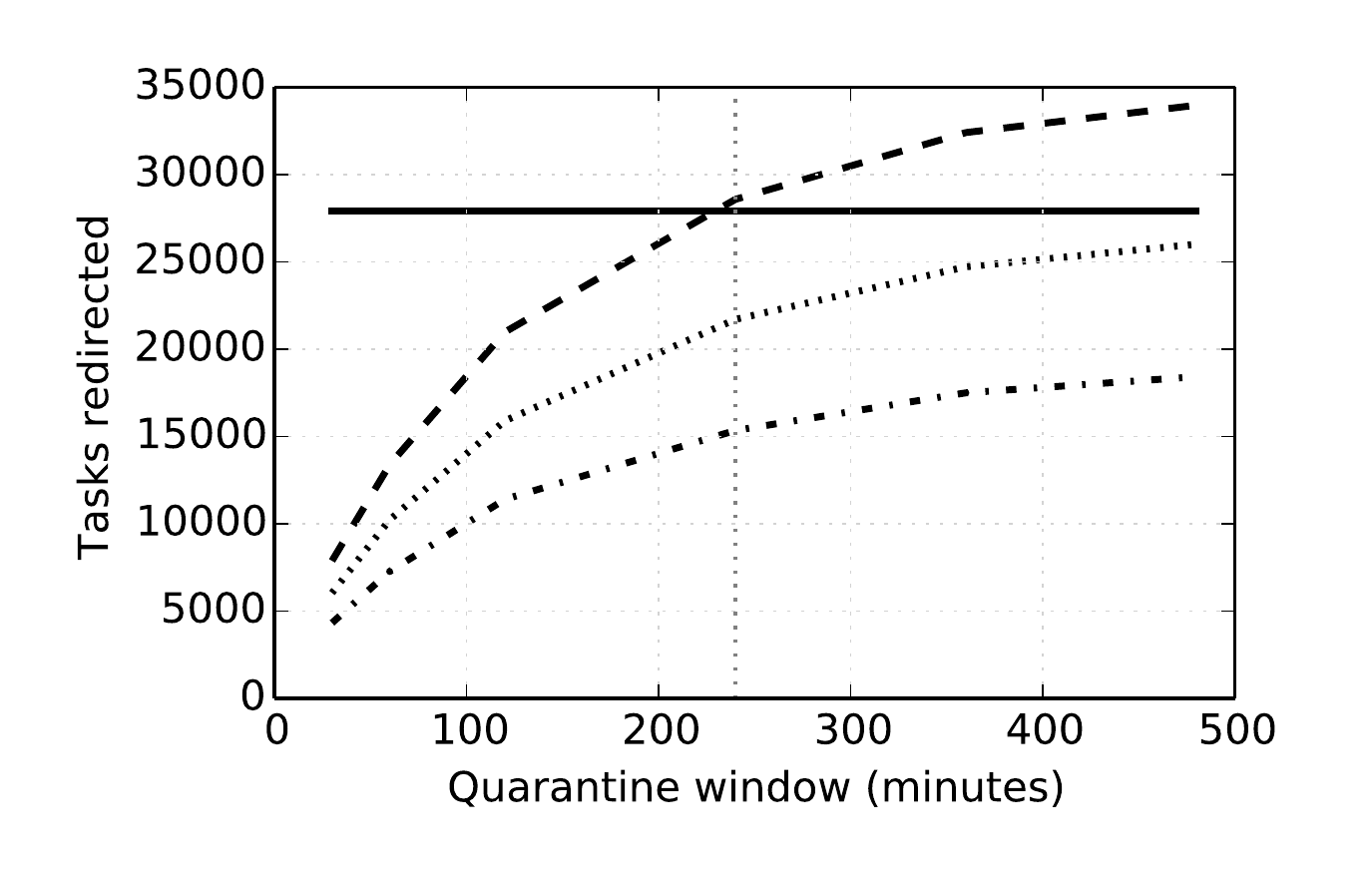}%
\label{fig_ttr15_c}}
\subfloat[CPU time redirected]{\includegraphics[width=2.7in]{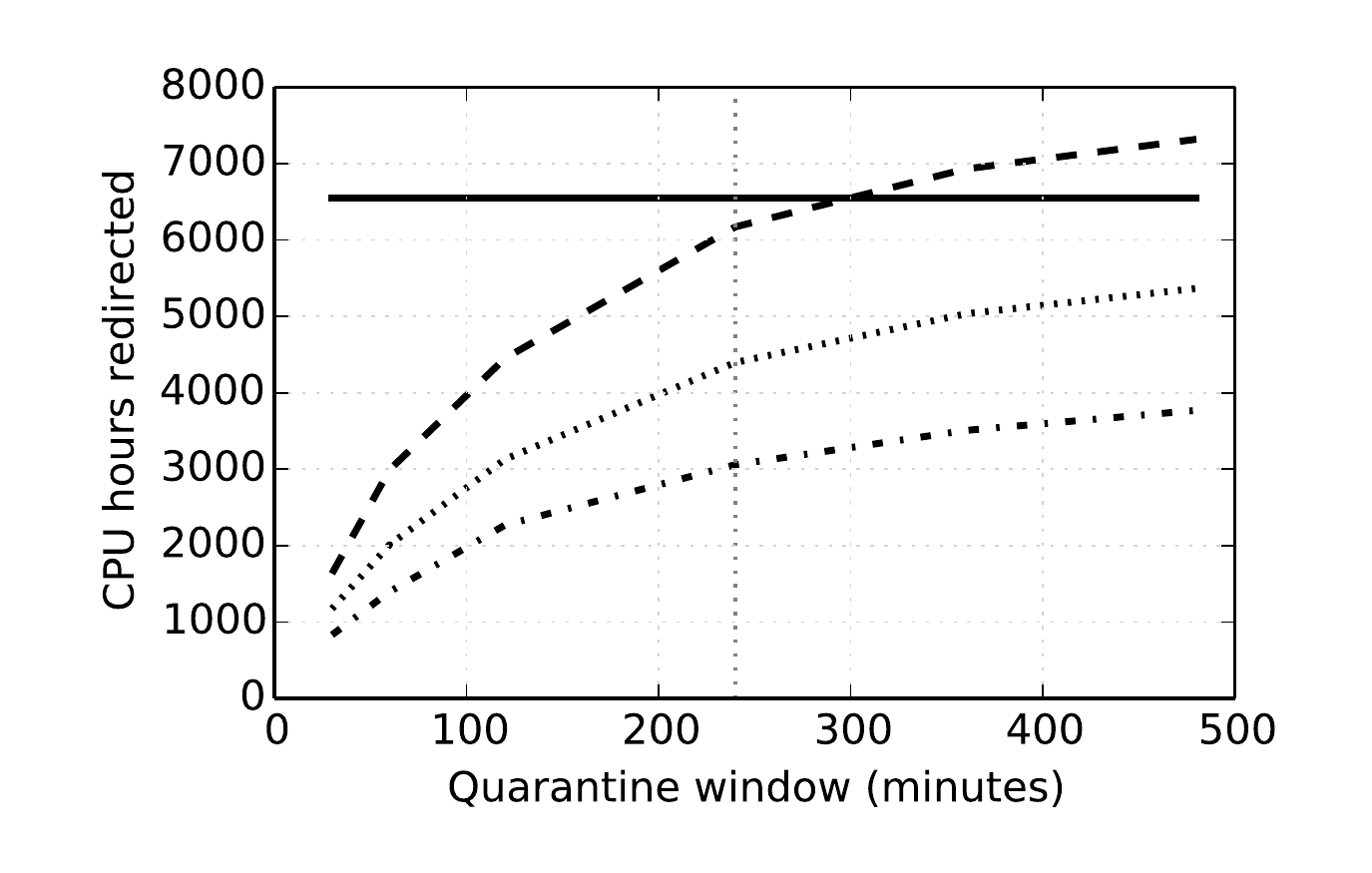}%
\label{fig_ttr15_d}}
\caption{Number of tasks and CPU hours redirected and recovered after prediction, compared to the baseline given by perfect prediction results.}
\label{fig_saved}
\end{figure*}

While a node is in quarantine, all tasks that would have otherwise run on that node need to be \emph{redirected}. Among redirected tasks, some would have finished before the node failure, others would have been \emph{interrupted}. We call the latter \emph{recovered} tasks, since their interruption was avoided by proactively redirecting them. The aim of our proactive approach is to \emph{maximize} the number of \emph{recovered} tasks (the gain) while \emph{minimizing} the number of \emph{redirected} tasks (the cost). As we shall see later in this section, the two objectives are contradictory: the number of \emph{recovered} tasks grows as the number of \emph{redirected} tasks increases, so there is a tradeoff between the cost and the gain.

To provide a baseline for our results, we estimated the number of \emph{recovered} tasks if \emph{perfect prediction were possible 24 hours in advance} on our data.
Perfect prediction means that a failure alarm is triggered every 5 minutes, starting 24 hours before the failure. Hence, nodes will enter quarantine state 24 hours before their failure and remain in quarantine until they fail.
Since lead time for failure prediction is limited (for us it is 24 hours), not all interrupted tasks can be recovered (those that started before the first failure alarm). Table~\ref{table_baseline} shows the number of \emph{interrupted}, \emph{recovered} and \emph{redirected} tasks along with the respective CPU-hours, globally and according to scheduling class, assuming perfect prediction. Since tasks in lower scheduling classes are shorter, they are easier to recover since they can start after the failure alarms. In particular, 82.4\% of the interrupted tasks of scheduling class 0 and only 24.8\% of scheduling class 3 can be recovered by perfect prediction, corresponding to recovering 57.5\% and 12.3\%  CPU-hours, respectively.
As discussed earlier, for higher scheduling classes, task interruption does not necessarily mean that all used resources are wasted.  This makes estimating the exact resource wastage for high scheduling class tasks very difficult. Hence, in the following we limit our analysis to tasks from scheduling class 0, where all resources used by interrupted tasks can be considered wasted.

Taking Table~\ref{table_baseline} (i.e., perfect prediction) as a baseline, we evaluated the effect of our prediction method on running tasks.  When using our classifier model for predictions, we may have false alarms resulting in a node to be quarantined without reason, while other alarms may be missed, so the node may not be quarantined for an entire 24 hours before it fails. In order to study the tradeoff between cost and gain, we explored various quarantine windows and different False Positive Rates (FPR) that characterize our classification.  Ideally, our method should be able to recover a large percentage of the baseline \emph{recovered} tasks, while not exceeding the baseline \emph{redirected} tasks significantly. 

Fig.~\ref{fig_saved} shows the number of \emph{recovered} and \emph{redirected} tasks for our method when limited to scheduling class 0, along with the corresponding CPU-hours. The number of \emph{recovered} tasks increases with increasing the FPR and the quarantine windows, however at the cost of increasing the total number of \emph{redirected} tasks. A quarantine window size of 4 hours and FPR~$=0.2$ appears to provide a good compromise between \emph{recovered} and \emph{redirected} tasks, with about 48\% of baseline tasks and 50\% of baseline CPU time \emph{recovered} and 102\% of baseline tasks and 96\% baseline CPU hours \emph{redirected}.
Hence our technique can have a significant impact on running tasks with respect to the 
baseline.
A more sophisticated strategy can further enhance both the baseline and our classification results.

}
\section{Discussion}\label{discussion}

We now consider how our predictive model can be the basis for a data-driven ``Autonomics 2.0'' controller to be deployed in data centers. In such a scenario, 
both model building and model updating 
would have to happen on-line on data that is being streamed from various sources.
We outline some of the changes required by our model workflow to allow on-line use. 

To compute on-line the features necessary for model building, log data can be collected in a BigQuery table using the streaming API. As data is being streamed, features have to be computed at 5 minute intervals. Both basic and aggregated features (averages, standard deviations, coefficients of variation and correlations) have to be computed, but only for the last time window (previous time windows are already stored in a dedicated table). Basic features are straightforward to compute requiring negligible running time since they can be computed using accumulators as the events come in.  Aggregated features can be computed in parallel since they are independent of each other. In our experiments, correlation computation was most time consuming, with an average time to compute one correlation feature over the longest time window taking 1489.2 seconds for all values over the 29 days (Table \ref{table_times}). For computing a single value (for the newly streamed data), the time required should be on average under 0.2 seconds. This estimate is based on a linear dependence between the number of time windows and computation time and offers an upper bound for the time required. If this stage is performed in parallel on BigQuery, this value would also be the average time to compute all 126 correlation features (each feature can be computed independently so speedup would be linear). 

In terms of dollar costs, we expect figures similar to those during our tests --- about 70 USD per day for storage and analysis. To this, the streaming costs would have to be added --- currently 1 cent per 200MB. For our system, the original raw data is about 200GB for all 29 days, so this would translate to approximately 7GB of data streamed every day for a system of similar size, resulting in about 35 cents of additional cost per day.  At all times, only the last 12 days of features need to be stored, which keeps data size relatively low. In our analysis, for all 29 days, the final feature table requires 295GB of BigQuery storage, so 12 days would amount to about 122GB of data.

When a new model has to be trained (e.g., once a day), all necessary features are already computed. One can use an infrastructure like the Google Compute Engine to train the model, which would eliminate the need to download the data and would allow for training of the individual classifiers of the ensemble in parallel. In our tests, the entire ensemble took under 9 hours to train, with each RF requiring at most 3 minutes. Again, since each classifier is independent, training all classifiers in parallel would take under 3 minutes as well (provided one can use as many CPUs as there are RFs --- 420 in our study). Combining the classifiers requires a negligible amount of time. 

All in all, we expect the entire process of updating the model to take under 5 minutes if full parallelization is used both for feature computation and training. Application of the model on new data requires a negligible amount of time once features are available. This makes the method very practical for on-line use. Here we have described a cloud computing scenario, however, given the relatively limited computation and storage resources that are required, we believe that more modest clusters can also be used for monitoring, model updating and prediction. 
 
\section{Related work}\label{related}
The publication of the Google trace data has triggered a flurry of activity within the community including several with goals that are related to ours. Some of these provide general characterization and statistics about the workload and node state for the cluster \cite{Reiss2012,Reiss2012a,Liu2012} and identify high levels of heterogeneity and dynamism in the system, especially when compared to grid workloads \cite{Di2012}. User profiles \cite{Abdul-Rahman2014} and task usage shapes \cite{Zhang2011} have also been characterized for this cluster. Other studies have applied clustering techniques for workload characterization, either in terms of jobs and resources \cite{Mishra2010,Wang2011} or placement constraints \cite{sharma2011}, with the aim to synthesize new traces. \revised{Wasted resources due to the priority-based eviction mechanism were evaluated in \cite{Rosa2015}, were significant resources were shown to be used by tasks that do not complete successfully, as we have also seen in our analysis of tasks interrupted by failures.}

A different class of studies address validation of various workload management algorithms. Examples include \cite{Iglesias2014} where the trace is used to evaluate consolidation strategies, \cite{Caglar2014,Breitgand2014} where  over-committing (overbooking) is validated, \cite{Zhang2014} that takes heterogeneity into account to perform provisioning or \cite{Di2013a} investigating checkpointing algorithms. 

System modeling and prediction studies using the Google trace data are far fewer than those aimed at characterization or validation. An early attempt at system modeling based on this trace validates an event-based simulator using workload parameters extracted from the data, with good performance in simulating job status and overall system load~\cite{Balliu2014,Balliu2015}. Host load prediction using a Bayesian classifier was analyzed in~\cite{Di2012a}. Using CPU and RAM history, the mean load in a future time window is predicted by dividing possible load levels into 50 discrete states. \revised{Job failure prediction and mitigation is attempted in \cite{Rosa2015b}, with 27\% of wasted computational time saved by their method.} Here we investigate predicting \emph{node failures} for this cluster, which to our knowledge has not been attempted before.
 
Failure prediction in general has been an active research area, with a comprehensive review~\cite{Salfner2010} summarizing several methods of failure prediction in single machines, clusters, application servers, file systems, hard drives, email servers and clients, etc., dividing them into failure tracking, symptom monitoring or error reporting. The method introduced here falls into the symptom monitoring category, however elements of failure tracking and error reporting are also present through features like number of recent failures and job failure events. 

More recent studies concentrate on larger scale distributed systems such as HPC or clouds. For failure tracking methods, an important resource is the failure trace archive \cite{Javadi2013}, a repository for failure logs and an associated toolkit that enables integrated characterization of failures, such as distributions of inter-event times. Job failure in a cloud setting has been analyzed in \cite{samak2012}. The naive Bayes classifier is used to obtain a probability of failure based on the job type and host name. This is applied to traces from Amazon EC2 running several scientific applications. The method reaches different performances on jobs from three different application settings, with FNs of 4\%, 12\% and 16\% of total data points and corresponding FPs of 0\%, 4\% and 10\% of total data points. This corresponds approximately to FPR of 0\%, 5.7\% and 16.3\%, and 
TPR of 86.6\%, 61.2\% and 58.9\%. The performance we obtained with our method is within a similar range for most benchmarks, although we never reach their best performance. However, we are predicting node failures rather than job failures.  \revised{A recent study of the Blue Waters HPC installation at Argonne National Laboratory predicted failures with 60\% TPR and 85\% precision \cite{Gainaru2014}, again representing performance similar to ours.}

A comparison of different classification tools for failure prediction in an IBM Blue Gene/L cluster is given in \cite{Liang2007}.
In this work, Reliability, Availability and Serviceability (RAS) events are analyzed using SVMs, neural networks, rule based classifiers and a custom nearest neighbor algorithm, in an effort to predict whether different event categories will appear. The custom nearest neighbor algorithm outperforms the others reaching 50\% precision and 80\% TPR.  A similar analysis was also performed for a Blue Gene/Q cluster~\cite{Dudko2012}. The best performance was again achieved by the nearest neighbor classifier (10\% FPR, 20\% TPR). They never evaluated the Random Forest or ensemble algorithms.

In \cite{Qiang2013} an anomaly detection algorithm for cloud computing is introduced.
It employs Principal Component Analysis  and selects the most relevant principal components for each failure type. They find that higher order components exhibit higher correlation with errors. Using a threshold on these principal components, they identify data points outside the normal range. They study four types of failures: CPU-related, memory-related, disk-related  and network-related faults, in a controlled in-house system with fault injection and obtain very high performance, 91.4\% TPR at 3.7\% FPR. On a production trace (the same Google trace we are using) they predict task failures at 81.5\% TPR and 27\% FPR (at 5\% FPR, TPR is down to about 40\%). In our case, we studied the same trace but looking at node failures as opposed to task failures, and obtained TRP values between 27\% and 88\% at 5\% FPR.

All above-mentioned failure prediction studies concentrate on types of failures or systems different from ours and obtain variable results. In all cases, our predictions compare well with prior studies, with our best result being better than most.

\section{Conclusions}\label{conclusions}

We have presented study of node failure models for data centers based on log data from a Google cluster.
Such models will be an integral part of a new generation of ``Autonomics 2.0'' architectures.
Our results confirm that models with sufficient predictive powers can indeed be built based on data found in typical logs and that they can form the basis of an effective autonomic manager that is data driven, predictive and proactive.

Model feature extraction from the raw data was performed using BigQuery, the big data cloud platform from Google that supports SQL-like queries. A large number of features were generated and an ensemble classifier was trained on log data for 10 days and tested on the following non-overlapping day. The length of the trace allowed repeating this process 15 times producing 15 benchmark datasets, with the last day in each dataset being used for testing. 

The BigQuery platform was extremely useful for obtaining the features from log data. Although limits were found for \textsc{join} and \textsc{group by} statements, these were circumvented by creating intermediate tables, which at times contained more than 12TB of data.
Even so, features were obtained 
in a reasonable amount of time
with overall cost for the entire analysis processing a one-month log coming in at under 2,000 USD\footnote{Based on current Google BigQuery pricing.},
resulting in a daily cost under 70 USD. 

Classification performance varied from one benchmark to another, with Area-Under-the-ROC curve measure varying between 0.76 and 0.97 while Area-Under-the-Precision-Recall curve measure varying between 0.38 and 0.87. This corresponded to true positive rates in the range  27\%-88\% and precision values between 50\% and 72\% at a false positive rate of 5\%. In other words, this means that in the worst case, we were able to identify 27\% of failures, while if a data point was classified as a failure, we could have 50\% confidence that we were looking at a real failure. For the best case, we were able to identify almost 90\% of failures and 72\% of instances classified as failures corresponded to real failures. All this, at the cost of having a 5\% false alarm rate.

Although not perfect, our predictions achieve good performance levels.
Results could be improved by changing the subsampling procedure. Here, only a subset of the \textsc{safe} data was used due to the large number of data points in this class, and a random sample was extracted from this subset when training each classifier in the ensemble when in fact one could subsample every time from the full set. However, this would require greater computational resources for training, since a single workstation cannot process 300 GB of data at a time. Training times could be reduced through parallelization, since the problem is embarrassingly parallel (each classifier in the ensemble can be trained independently from the others). These improvements will be pursued in the future. Introduction of additional features will also be explored to take into account in a more explicit manner the interaction between nodes. BigQuery will be used to extract these interactions and build networks to model them. Changes in the properties of these networks over time could provide important information on possible future failures.

The method presented here is suitable for on-line use. A new model can be trained every day using the last 12 days of logs. This is the scenario we simulated when we created the 15 test benchmarks. The model would be \emph{trained} with 10 days of data and \emph{tested} on the next non-overlapping day, exactly like in the benchmarks (Fig.~\ref{fig_xval}). Then, it would be \emph{applied} for one day to predict future failures. The next day a new model would be obtained from new data.
Each time, only the last 12 days of data would be used rather than increasing the amount of training data. This would account for the fact that the system itself and the workload can change over time, so old data may not be representative of current system behavior.
This would ensure that the model is up-to-date with the current system state.
Testing on one non-overlapping day
is required for live use for two reasons.
First, part of the test data is used to build the ensemble (prediction-weighted voting). Secondly, the TPR and precision values on test data can help system administrators make decisions on the \emph{criticality of the predicted failure} when the model is applied. \revised{Here, we explored how simply putting the nodes that are predicted to fail under quarantine can help recover many of the interrupted tasks and save some of the wasted resources.
More sophisticated strategies can be based on 
the criticality of predicted failures. 
}

\section{Acknowledgements}
BigQuery analysis was carried out through a generous Cloud Credits grant from Google. We are grateful to John Wilkes of Google for helpful discussions regarding the cluster trace data.

\bibliographystyle{plain}      
\bibliography{refs}   

\end{document}

%% file: datapoints.txt
1&41485&2055&9609&2055&9610\\

2&41005&2010&9408&2011&9408\\

3&41592&1638&9606&1638&9606\\

4&42347&1770&9597&1770&9598\\

5&42958&1909&9589&1909&9589\\

6&42862&1999&9913&2000&9914\\

7&41984&1787&9821&1787&9822\\

8&39953&1520&10424&1520&10424\\

9&37719&1665&10007&1666&10008\\

10&36818&1582&9462&1583&9463\\

11&35431&1999&9302&1999&9302\\

12&35978&3786&10409&3787&10410\\

13&35862&2114&9575&2114&9575\\

14&39426&1449&9609&1450&9610\\

15&40377&1284&9783&1285&9784\\